\documentclass[11pt,twoside]{article}

\usepackage{asp2006}
\usepackage{epsf}
\usepackage{psfig}
\usepackage{lscape}

\begin{document}
\title{Iron K Line Diagnostics in Active Galactic Nuclei}   
\author{Tahir Yaqoob}   
\affil{Johns Hopkins University, 3400 N. Charles St., Baltimore, MD~21218. 
NASA/GSFC, Code 662, Greenbelt Rd., Greenbelt, MD~20771, USA.}    

\begin{abstract} 
We discuss some topical issues related to the study of Fe~K emission
lines in Active Galactic Nuclei (AGNs). 
We show
remarkable agreement between {\it non-contemporaneous} ASCA and
{\it Chandra} grating data and
explain why
there has been terrible confusion about the ASCA and post-ASCA
results on the relativistic components of the Fe~K line emission.
We point out that in fact the {\it number of sources} (not the
percentage) that have been reported to
exhibit relativistic effects in the Fe~K line is now larger
than it was in the ASCA era.
Thus, the case for
{\it Constellation-X} as a probe of strong gravity is even
more compelling than it was a decade ago.
One of the primary goals of these studies is to establish the foundations
for future missions to map the spacetime metric around black holes.
A prerequisite first step 
is to measure the black-hole angular momentum
in a robust manner that does not rely on assumptions about the
accreting system. 
In addition, probing
the Fe~K lines out to high redshifts will pave
the way for
studying the accretion history and evolution of supermassive
black holes. However, we point out some issues
that need to be resolved,
pertaining to spin measurement and to the relativistic Fe~K
line emission found from AGN in deep X-ray surveys.
\end{abstract}

\section{Introduction}
Black holes have ``no hair'' and have only three measurable
parameters (not including the Hawking temperature):
charge, mass, and angular momentum. 
We do not yet know how to measure the charge. Measurements of black-hole
mass are well underway and constitute a field in itself and
we do not discuss it here (e.g. see Peterson \& Bentz 2006,
and references therein). That leaves black-hole
angular momentum, or spin. This is a property of the space-time metric,
and measurement of the spin would be a first step towards mapping
the metric. If one cannot measure the spin there is little
hope of mapping the metric in order to compare with the predictions
of general relativity (or other theories of gravity).

\begin{figure}[!thb]
\centerline{
\psfig{{file=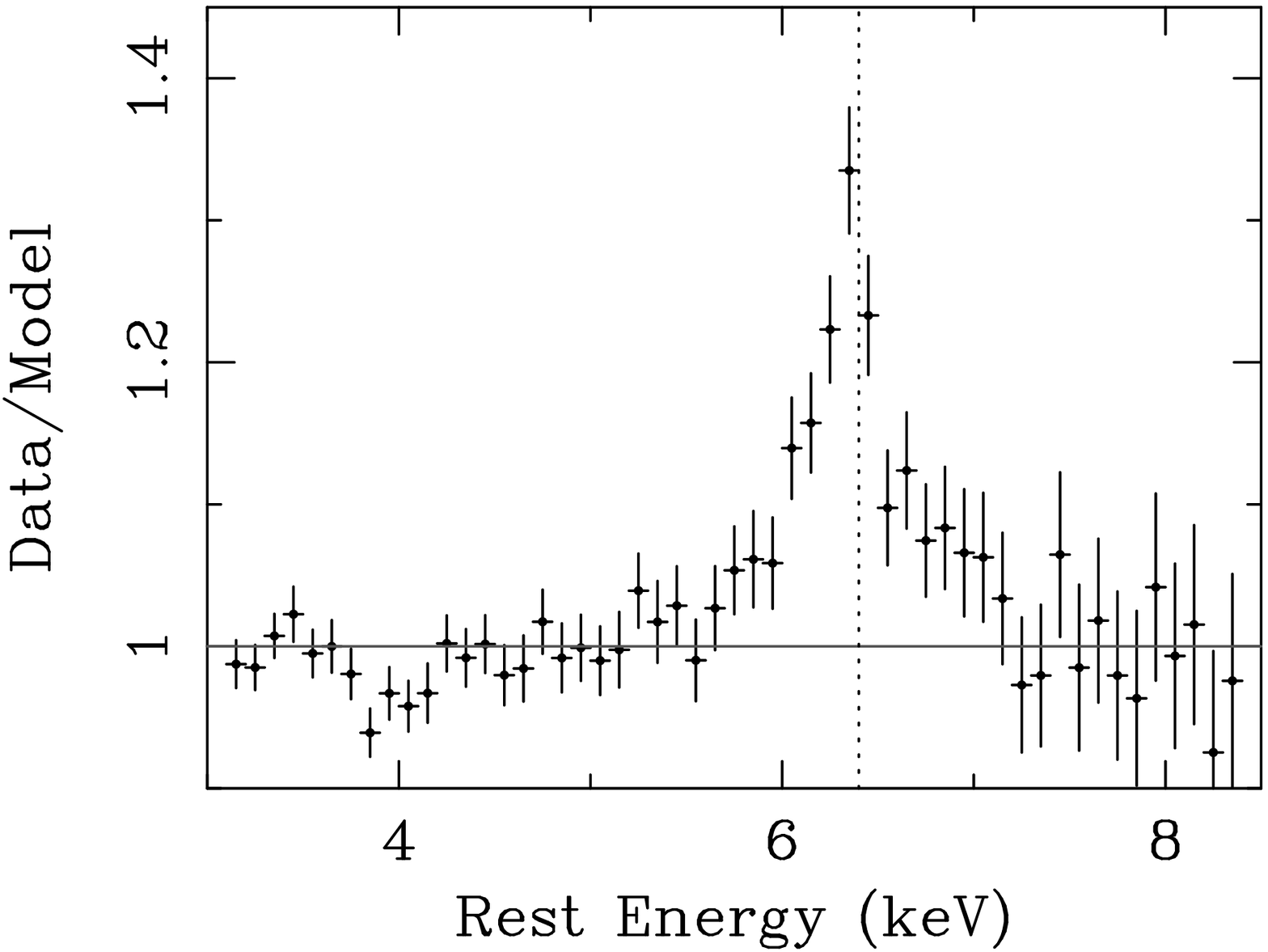,width=4.3cm,height=3.33cm}}
\psfig{file=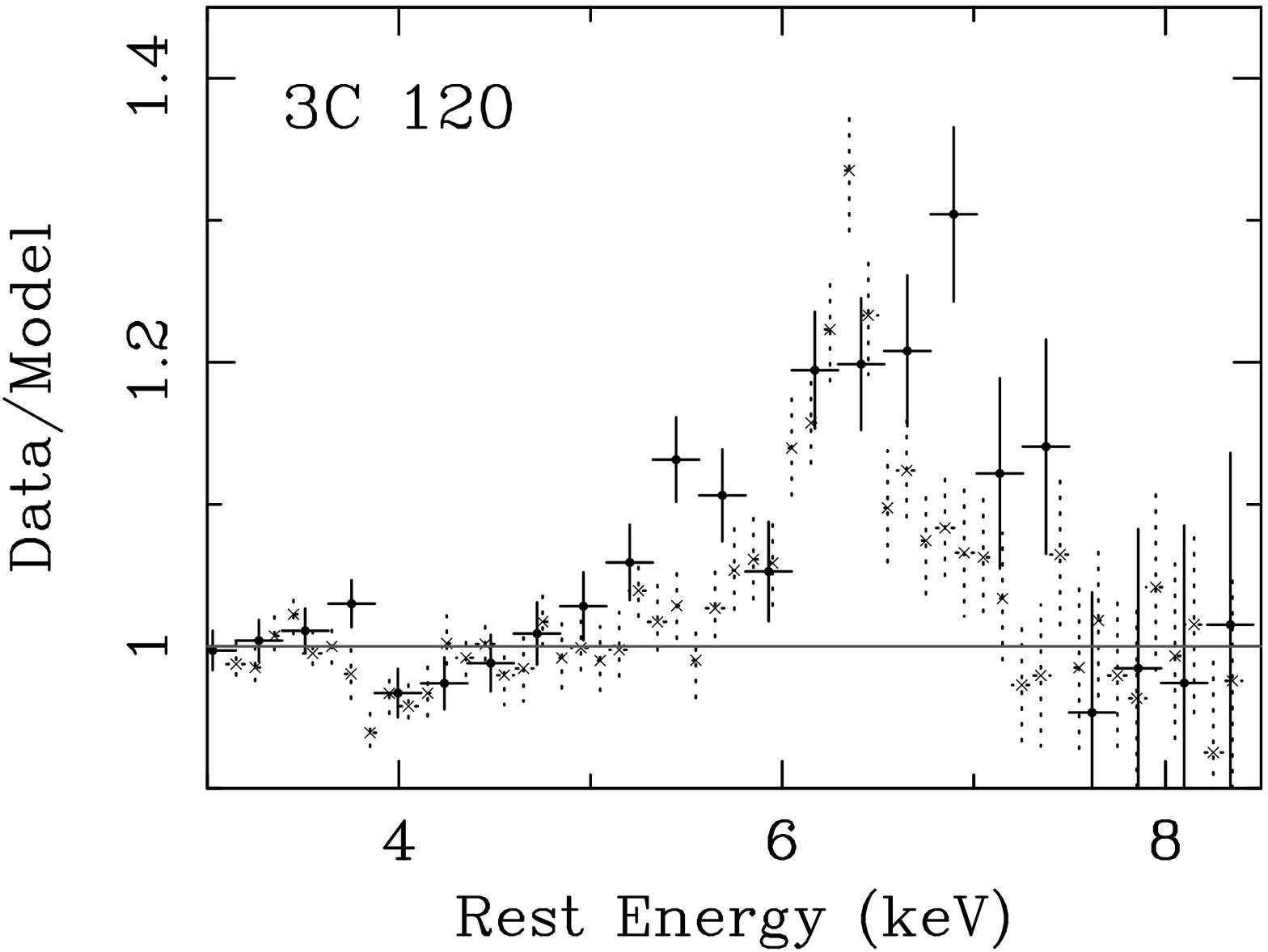,width=4.3cm,height=3.33cm}
\psfig{file=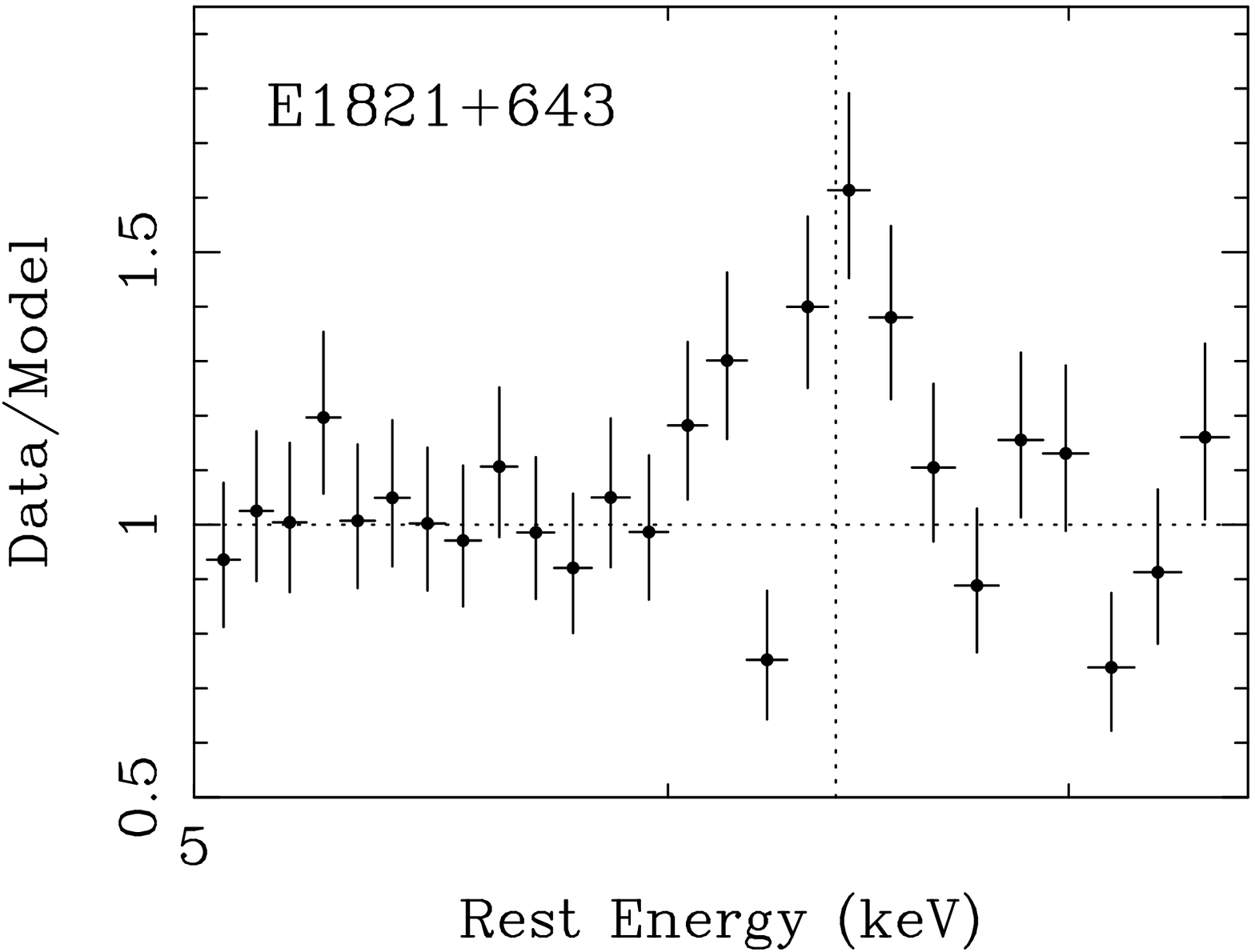,width=4.3cm,height=3.33cm}
}
\vspace{-2mm}
\caption{{\it Left:} (a) The mean ASCA Fe~K line profile (see Yaqoob et al. 2002).
{\it Middle:} (b) The ASCA (AO6) 
Fe~K line profile of 3C~120 (solid) compared
with (a). {\it Right:} (c) The highest luminosity, highest
redshift ($z=0.297$) radio-quiet AGN with a reported Fe~K line
($E~1821$+$643$). Note the redshifted absorption line in $E~1821$+$643$
(see Yaqoob \& Serlemitsos 2005).
}
\end{figure}

Currently, {\it detailed} Fe~K line spectroscopy is limited to
low-redshift AGNs, plus a handful of high-redshift objects.
Eventually we would like to be able to routinely measure
the black-hole spin for AGNs out to all observable redshifts.
The spin may be a function of some other physical
property of accreting black-hole systems that may also
be time-dependent and evolving.
For example, according to Thorne (1974), 
disk accretion increases the angular momentum of
a black-hole, reaching a 
maximal value of $a/M = 0.9982$ (but see Gammie, Shapiro, \&
McKinney 2004).
Thus, the Fe~K line emission in AGN could eventually be used to
study accretion in a cosmological context, in
particular the accretion history, and the evolution of accreting
black holes.

The Fe~K emission line
is also a probe of the physics and
structure of the accretion flow itself.
In particular, the inclination angle of the disk
with respect to the observer, the ionization state of the disk
(and its dependence on the source X-ray luminosity and 
accretion rate),
and its inner radius are important physical properties that
still remain largely elusive.
Finally, the narrow Fe~K line emission from more distant matter
beyond the accretion disk carries important information on
the structure and physical state of matter in the outer regions
of the AGN central engine, possibly originating from the
putative parsec-scale ``obscuring torus'' that is a key
component of standard AGN unification schemes.
Although relativistic Fe~K lines are observed in X-ray
binaries (e.g. see Miller 2006, and references therein) 
and do not suffer from the problem of
deconvolution from a distant-matter Fe~K line, in this work
we focus on AGNs.

\section{Did the ASCA Broad Fe~K Lines Go Away?}

Since the first reports of the observation of relativistically
broadened Fe~K emission lines in AGN, 
many studies have been done of the same and additional
sources using 
{\it BeppoSAX} (e.g. Perola et al. 2002), 
{\it XMM-Newton} (e.g. Reynolds \& Nowak 2003;
Page et al. 2004; Porquet et al. 2004; 
Jim\'{e}nez-Bail\'{o}n et al. 2005; 
Fabian and Miniutti 2005; Jiang, Wang \& Wang 2006 and references
therein), and Suzaku (e.g. Miniutti et al. 2006; Reeves et al. 2006;
Yaqoob et al. 2006).
There is a myth that
the post-ASCA results are somehow inconsistent with ASCA,
and that the broad Fe~K lines are less
common than ASCA had found. 
Lubi\'{n}ski \& Zdziarski (2001) began the important process
of re-examining the ASCA data.
Changes in the ASCA calibration
were shown to have a negligible impact on the 
Fe~K line profiles
(Yaqoob et al. 2002), yet the myth and confusion 
still prevail.

The myth arose because,
given the currently available data
(ASCA and post-ASCA), the ``percentage of
AGN with a relativistically broadened
Fe~K line'' is a model-dependent quantity
that also depends on the criterion used to define ``broad''.
Despite many ASCA results appearing after
Nandra et al. (1997 -- hereafter, N97), it is this paper
that the confusion is centered around.
The claim in N97 was that fourteen out of eighteen type~1
AGN have a resolved Fe~K line. {\it ``Resolved by ASCA''} does not
mean that all fourteen sources have an enormous red wing on  
the Fe~K line as does MCG~$-$6$-$30$-$15. 
The lines were parameterized
by Gaussian models, the data were fitted between 3--10~keV
only, and the broad Gaussian
could in part be modeling complexity in the continuum (e.g.
due to ionized absorption).
Of these fourteen AGN
at least four should not have been counted.
NGC~7469 has a 68\% confidence error (table~3 in N97) that does not
exclude the line being unresolved. Interestingly, the
line profile has a strong blue wing and virtually no red wing.
For NGC~6814, the signal-to-noise is so low that no actual
line is apparent in the spectrum and the Gaussian component
was in fact modeling the continuum. The signal-to-noise in the
brighter Mkn~841 observation was also too low to distinguish line emission
from continuum emission - this can be seen by eye. The line
profile for NGC~5548 shows a small excess on the blue side of
the peak but, after accounting for the spectral resolution,
there appears to be no excess on the red side. We note that
excess emission blue-ward of the Fe~K line peak can be
caused by complexity in the continuum. For example, 
it can clearly be seen from the photoionization models 
applied to MCG~$-$6$-$30$-$15 
by Lee et al. (2001) that 
continuum curvature and edge features are apparent all the way up
to the Fe~K band and are not just restricted to below 3~keV.
Thus, we are left with ten {\it or less} possible broad lines in the
ASCA sample and this corresponds to 56\% {\it or less}. 
This is
{\it not} inconsistent with post-ASCA results.
Guainazzi, Bianchi, \& Dov\v{c}iak (2006) found
that $\sim 25-50\%$ of low-redshift AGN
show broad relativistic lines, the actual value depending on
sample selection (see also Nandra et al. 2006). 
Model-dependence of the broad-line
detections and measurements contributes additional
uncertainty.
If post-ASCA results are compared
with the ASCA results on a source-by-source basis,
as opposed to some sample property or mean line profile,
no problem arises.

Another factor is over-interpretation of mean Fe~K
line profiles from the ASCA sample. 
Figure~1a shows a later version of one of the
mean profiles in N97 (see Yaqoob et al. 2002
for details).
It excludes MCG~$-$6$-$30$-$15, NGC~4151, 
and NGC~6814 (the latter because
it is so dim). 
There is a red wing but it is weak and its relative strength
is sensitive to the continuum level.
However, the version of the profile that most people
associate with ASCA is the one that did not exclude the above
three sources and shows a much stronger red wing,
biased by MCG~$-$6$-$30$-$15 and NGC~4151. Although N97 showed
a profile excluding these latter two (that
is consistent with Figure~1a), it is not the version people
recall. However,
we emphasize that {\it there are now a larger number
of AGN with reported broad Fe~K lines than in the ASCA era}
even though that number as a fraction of a class
may have been revised. 
There are still strong candidates in the original ASCA sample.
Figure~1b shows the Fe~K line profile from an ASCA AO6 observation
of 3C~120 (N97 used a shorter, AO1 observation). 
These data show one of the broadest lines in the sample
(but this is not a unique interpretation).
A significant number of new broad Fe~K lines have been found 
with post-ASCA data.
Figure~1c shows the (broad) Fe~K line profile found in
the highest
luminosity, highest redshift quasar yet,
$E~1821+643$ (z=0.297, $L_{\rm 2-10 \ \rm keV}
\sim 3 \times 10^{45} \ \rm ergs \ s^{-1}$;
see Fang et al. 2002; Yaqoob \& Serlemitsos 2005).

\begin{figure}[!htb]
\centerline{\psfig{file=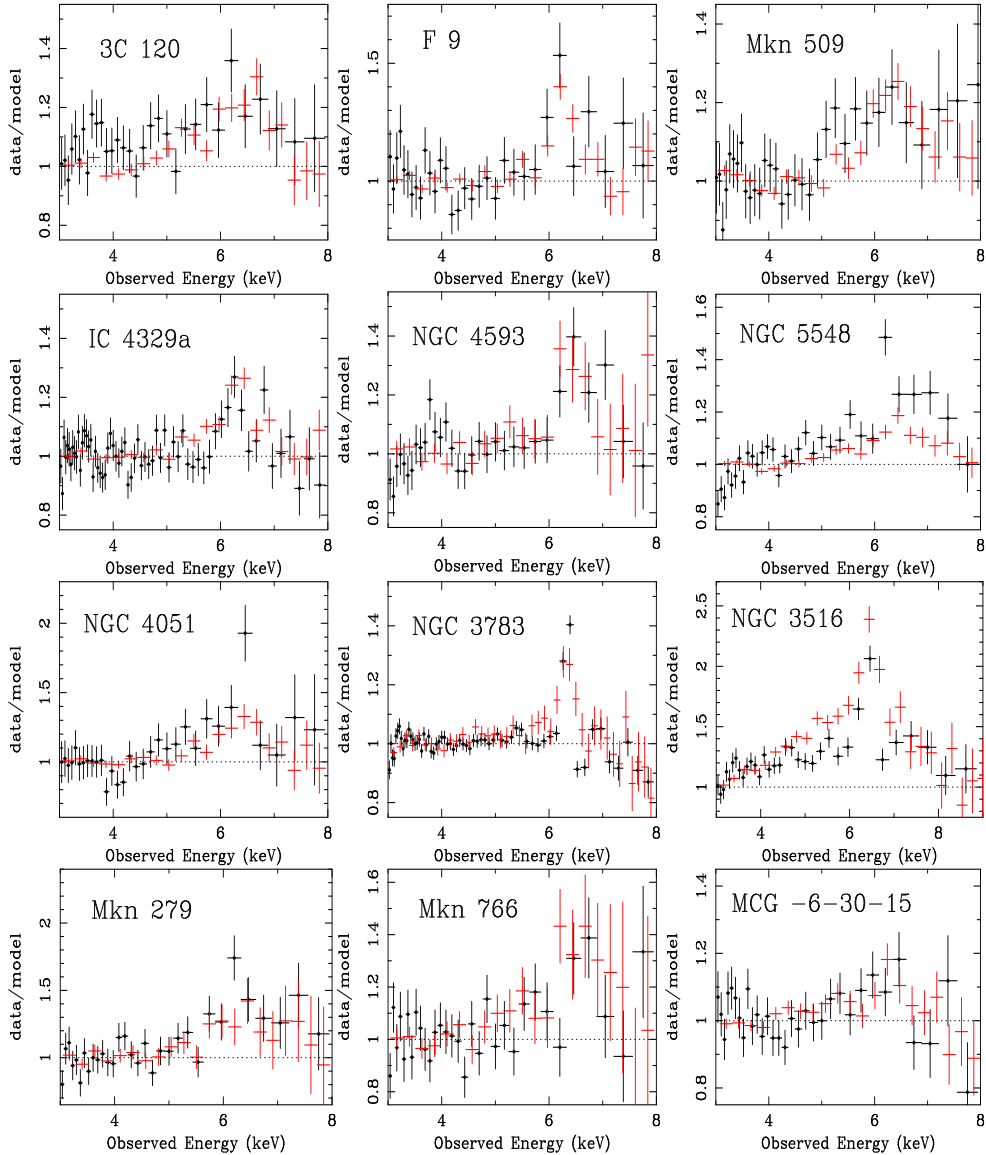,angle=0,width=13cm,height=15.3cm}
}
\vspace{-3mm}
\caption{The {\it Chandra} High Energy Grating (HEG) Fe~K line profiles
for twelve AGN in the Yaqoob \& Padmanabhan (2004) sample
(black) compared with 
non-contemporaneous ASCA data (red).
}
\end{figure}

Figure~2 shows heavily binned
{\it Chandra} 
High Energy Grating (HEG) data {\it directly compared to ASCA
data} for twelve type~1 AGN. The spectral resolution of the
HEG data is $\sim 1800 \rm \ km \ s^{-1}$ FWHM but here
we have binned it to approximately CCD resolution.
All except Mkn~279 were members of the N97 sample
and all are members of the {\it Chandra} HEG sample of
Yaqoob \& Padmanabhan (2004) in which
the Fe~K line core was studied. In cases 
where there were multiple ASCA observations,
only the highest signal-to-noise observation is shown
in Figure~2. The data/model ratios in Figure~2 were made
by fitting the 3--10~keV band with a simple
power law continuum,
but excluding the 5--7~keV data.
What is remarkable about Figure~2 is that even though the
ASCA and {\it Chandra} observations were {\it not} 
contemporaneous (typically separated by years), the
apparent ASCA and {\it Chandra} Fe~K line profiles 
are largely consistent with each other, except for
some energy regions in a few sources.
{\it The ASCA data were not flawed}. Although most of the Fe~K
line profiles in Figure~2 appear to be broad, this does not of course
prove that broad Fe~K lines are present because the 
continuum is not modeled. One needs to model the continuum
rigorously in order to determine any conclusions about the
Fe~K line parameters. Certainly,
however one chooses to model the ASCA data, the
{\it Chandra} data can in general
be fitted well with that same model.

\section{The Deconvolution Problem}

In general, that part of the Fe~K line emission
from a relativistic accretion disk that is
near the line rest-frame energy
(e.g. from the outer regions and/or line-emission observed
at small inclination angles to the disk normal)
is degenerate with line emission from distant matter
(e.g. see Weaver \& Reynolds 1998).
MCG~$-$6$-$30$-$15 is one of the exceptions, having a
narrow Fe~K line that is relatively weak.
Although the {\it Chandra} HEG has the best spectral
resolution, it is challenged by a small effective area. Some progress
in decoupling the broad and narrow Fe~K
line {\it intensities}
has been made with CCD data, but the narrow line is then 
even harder to resolve
(e.g. Reeves et al. 2006; Yaqoob et al. 2006).

The origin of the distant-matter Fe~K emission line remains
elusive. The peak line energies can be measured extremely well
and strong clustering around 6.4~keV 
indicates that the Fe is not highly ionized (e.g. 
Sulentic et al. 1998; Yaqoob \& Padmanabhan
2004). If the Fe~K$\beta$ line is detected with a good signal-to-noise 
ratio, the ionization state of Fe can be determined to
a very high precision due to the redundancy of information
(e.g. see Yaqoob et al. 2006).
The Fe~K line core widths measured by even the {\it Chandra}
gratings may still be affected by a contribution from any
underlying broad Fe~K line emission. Nevertheless,
in specific cases where one can show that the
Fe~K line width is {\it less} than the
optical, BLR line
width (such as that of $H\beta$) 
with a high statistical significance,
one can deduce that the distant-matter line in such sources
originates in matter farther out than the BLR.
Nandra (2006) illustrates some key examples, which are actually
more important than the lack of a correlation between the Fe~K and 
$H\beta$ line widths.
It is often argued that
a Compton-thin origin for the distant-matter Fe~K line is
ruled out because the EW of the line is too  
large but this argument assumes a time-steady situation over
many years. However, in any individual case it is not usually possible
to rule out continuum-line time delays as the cause for an
artificially high EW. Higher effective area 
as well as spectral resolution is needed to
unambiguously measure or rule out a ``Compton-shoulder''
in order to further constrain the origin.

\section{Disk Ionization and Other Suppressors of Relativistic Lines}

When a broad Fe~K emission line is detected,
given the current typical signal-to-noise ratio
even of {\it XMM-Newton} and Suzaku data, it is not generally possible
to robustly constrain the rest-frame energy of the line (and
therefore the ionization state of the relativistic line-emitting matter).
This is because there is 
degeneracy between the line rest energy and the
disk inclination angle and, to some extent, the line radial emissivity
function. The notion that the broad relativistic Fe~K lines 
originate predominantly in cold, neutral matter is another myth.
When disk-line plus Gaussian
models are fitted to data, the disk-line rest energy
generally has to be {\it assumed} 
and cannot be {\it derived}. 

There are notable cases where a peak in the Fe~K complex
is clearly higher than 6.4~keV and indicative of highly ionized
Fe. Obviously, for such line energy constraints to be measurable,
these emission lines that are unambiguously from ionized Fe are never
too broad -- if they were they could be modeled with disk lines
from neutral Fe with parameters that make the apparent line
centroid shift to higher energies.
In other cases, {\it XMM-Newton} or {\it Chandra} (grating) data reveal
that what was previously thought to be a single, broad line is 
in fact composed of multiple narrower lines, involving
multiple ionization stages of Fe (e.g. see Bianchi et al. 2005,
and references therein).  These narrow lines 
do not necessarily originate from distant matter,
based on variability and/or redshifting (e.g. see 
Turner et al. 2002; Petrucci et al. 2002; Yaqoob et al. 2003).
It also appears to be the case that
Fe~K emission lines from highly ionized Fe are more likely to
be found in NLS1s (Dewangan 2002),
which have steeper hard X-ray continua
than ``regular'' Seyfert~1 galaxies.

The fact that there are a significant number of radio-quiet,
low-luminosity AGN that have no detected broad Fe~K line but
have sufficient signal-to-noise to place tight upper limits on the
EW of such a line (e.g. Guainazzi et al. 2006 and references
therein) implies that there is an important factor missing in
our understanding. The explanation often invoked is that the
accretion disk is truncated so that the Fe~K line emission
does not extend down to small enough radii to cause sufficient
line broadening (e.g. see M\"{u}ller \& Camenzind 2004;
Matt et al. 2005,
and references therein). The truncation may be real or apparent (for
example if the Fe in the innermost region of the disk is 
completely ionized). Either way, this does not address the question
of what the fundamental driver is that causes the apparent
truncation.
The answer may not be simple because there are examples of
radio-quiet high-luminosity AGN that {\it do} have broad Fe~K lines
(see \S\ref{highzee}), yet ionized disks are usually invoked to
explain the lack of broad Fe~K lines in high-luminosity AGN. 
An effect that has not received much attention is that
if a broad Fe~K line emitted by a disk encounters a hot corona
with sufficient optical depth on the way to the observer,
it may be broadened so much that it may be undetectable against the
continuum. 
We emphasize that in order to obtain a complete
understanding of the broad Fe~K lines that {\it are} 
detected, {\it it will
be just as important for future missions such as Constellation-X
to observe the AGN in which no broad line
is currently detected}.

\section{High Luminosity and High Redshift}
\label{highzee}

The Fe~K emission lines 
(broad or narrow) in AGN become scarce at high luminosity
($L_{\rm 2-10 \ keV} > 10^{44} \ \rm ergs \ s^{-1}$ or
so), and at redshifts higher than $\sim 0.1$ or so.
The apparent anti-correlation of the Fe~K
emission line EW with the X-ray continuum luminosity 
has been dubbed the ``X-ray Baldwin
effect'' (Iwasawa \& Taniguchi 1993) and has been revisited many
times (e.g. Jiang et al. 2006, and references therein). 
However, both the measurement errors for the EW,
and the scatter in the correlation are 
large. Also, the highest luminosity AGN that have had pointed
X-ray observations tend to be mostly radio loud.
The presence of strong Fe~K lines in the
high luminosity radio-quiet quasars E~1821$+$643 
(e.g. Fang et al. 2002; Yaqoob
\& Serlemitsos 
2005) and Q0056$-$363 (Matt et al. 2005) destroys the X-ray
Baldwin effect, at least for radio-quiet AGN. 
E~1821$+$643 ($z=0.297$)
has $L_{\rm 2-10 \ keV} \sim 3 \times 10^{45} \rm \ ergs \ s^{-1}$
so it is the 
highest redshift, highest luminosity individual AGN known that has
a broad Fe~K emission line (see Figure~1c).
In comparison, Q0056$-$363 ($z=0.162$) has
$L_{\rm 2-10 \ keV} \sim 2 \times 10^{44} \rm \ ergs \ s^{-1}$.
Thus, it appears that the X-ray Baldwin effect may only be telling
us that radio-loud AGN have weak or no Fe~K line emission. For radio-quiet
AGN there appears to be no relation between the EW of the Fe~K line
and X-ray continuum luminosity. The EW of the Fe~K line in E~1821$+$643
is $209^{+51}_{-57}$~eV (Yaqoob \& Serlemitsos 2005) which is
comparable to (actually larger than) the typical EW for AGN with luminosities
at the low end of 
the range, $L_{\rm 2-10 \ keV} \sim 10^{42}  \rm \ ergs \ s^{-1}$
(e.g. see Page et al. 2004).

\section{The Fe~K Line in the Cosmological Context}

\begin{figure}[!htb]
\centerline{
\psfig{file=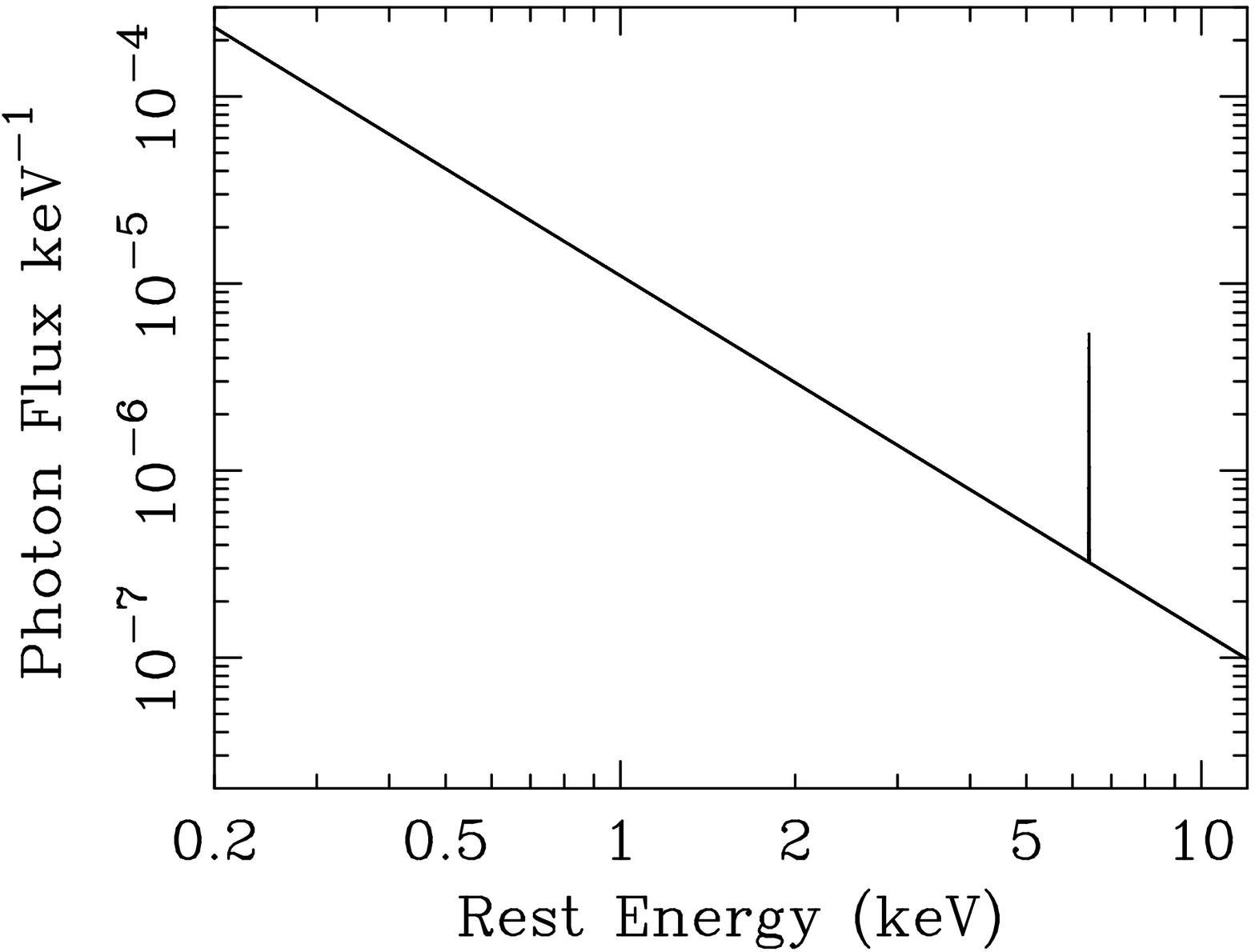,width=4.3cm,height=3.33cm}
\psfig{file=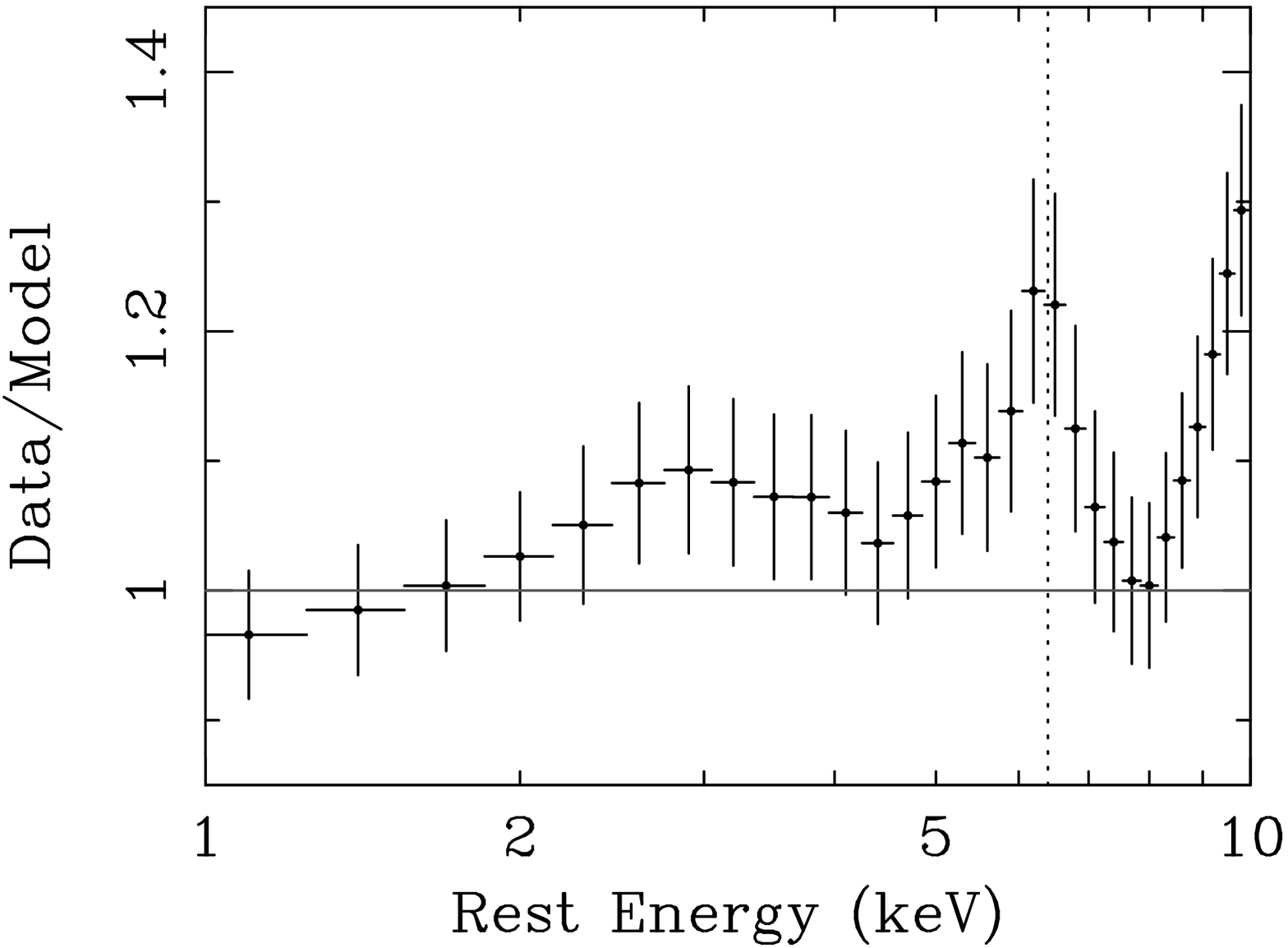,width=4.3cm,height=3.33cm}
\psfig{file=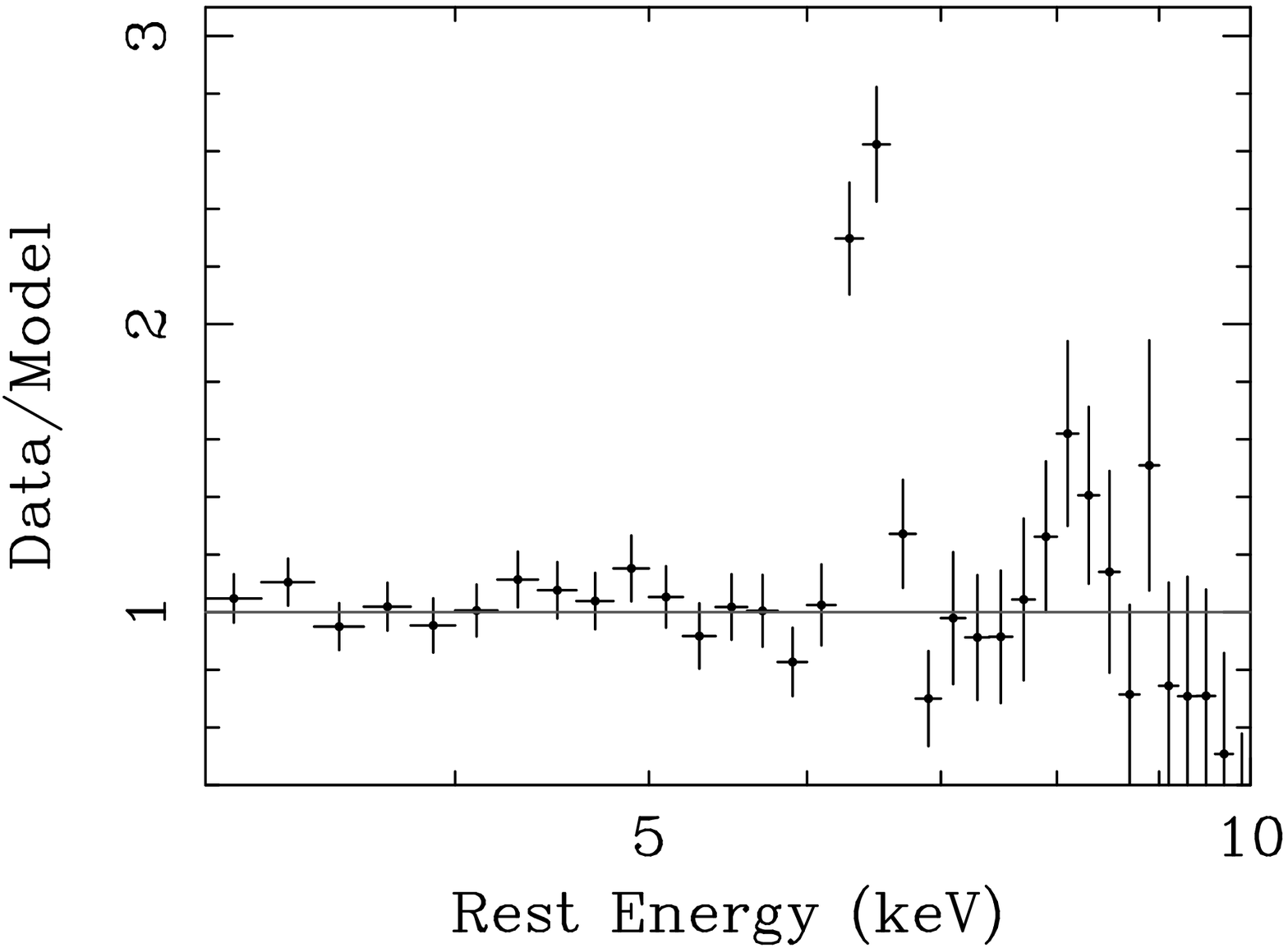,width=4.3cm,height=3.33cm}
}
\caption{Monte Carlo simulations of {\it XMM-Newton} spectra
of 200 AGN with a range in cosmological redshift. {\it Left:} (a) 
Input spectra with a narrow, unresolved Fe~K emission line.
{\it Middle:} (b) Data to power-law model ratio for mean spectrum
made using conventional methods: the spectral features are
artifacts of the averaging procedure. 
{\it Right:} (c) As (b) but using the new averaging procedure
in Yaqoob (2006). The narrow Fe~K line is recovered.  
}    
\end{figure}

The finding of broad relativistic Fe~K lines in the 
spectra of high-redshift sources found in deep X-ray surveys
would represent an extremely important milestone because
one could then use them to
study the history and evolution of accreting
black holes. Streblyanska et al. (2005)
have presented evidence
for such from the summed spectra of both type~1 and type~2 AGN
found in the Lockman Hole with {\it XMM-Newton} (each source
was too weak to yield line detection in any individual source).
The peak redshifts were $\sim 1.7$ for the type~1 AGN and
$\sim 0.7$ for the type~2 AGN respectively.
However, there is a fundamental problem with summing 
low signal-to-noise spectra over a range of redshifts.
This can introduce artificial features into the mean spectrum
that look just like relativistic line broadening and
a hard tail, mimicking a Compton reflection continuum (see Yaqoob 2006).
For example, Figure~3 shows the result of {\it XMM-Newton}
Monte Carlo simulations 
of 200 type~1 AGN with a redshift distribution similar to that of the
Streblyanska et al. (2005) sample, but with input spectra that were
simple power laws, with only {\it narrow, unresolved}, Fe~K emission 
lines. A new averaging method is given in
Yaqoob (2006) and Figure~3c shows that it successfully
recovers the input narrow line. The new method should now
be applied to real data. Brusa, Gilli, \& Comastri (2005)
studied Fe~K lines from {\it Chandra} deep field
sources, but summed the data in the observed frame
so they could not measure intrinsic line widths.

\section{Black-Hole Angular Momentum}

\begin{figure}[!htb]
\centerline{
\psfig{file=yaqoob_fig4a.eps,width=4.3cm,height=3.33cm,angle=270}
\psfig{file=yaqoob_fig4b.eps,width=4.3cm,height=3.4cm}
\psfig{file=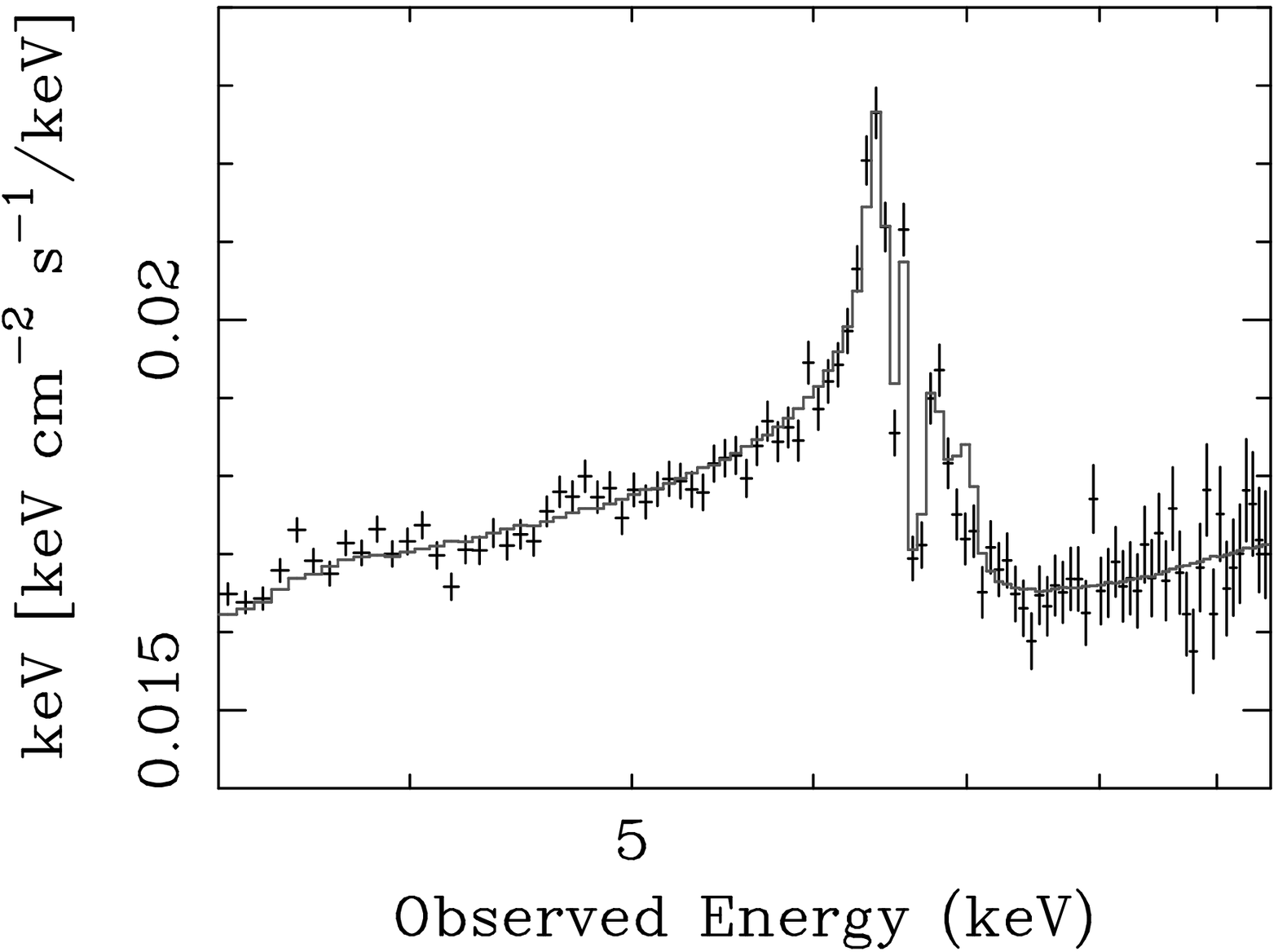,width=4.3cm,height=3.33cm}
}
\caption{{\it Left:} (a) The Fe~K line
profile at $6r_{g}$ with
$a/M=0$ (solid) and $a/M=1$ (dotted).
{\it Middle:} (b) Confidence contours (68\%, 90\%, and 99\%)
of disk inclination angle ($\theta_{0}$)
versus $a/M$ obtained
from fitting MCG~$-$6$-$30$-$15 data (using
a $\sim 290$~ks {\it XMM-Newton} EPIC pn spectrum). See Dovciak,
Karas, \& Yaqoob (2004) for details. 
{\it Right:} (c) A solution with Fe~K line emission around
a Schwarzschild black hole ($a/M=0$) fitted to the same data as (b),
{\it with no line emission from inside $6r_{g}$}.
}    
\end{figure}

A fundamental problem in measuring black-hole spin from the
Fe~K emission lines
is that the radial emissivity of the Fe~K line
from the accretion disk is {\it unknown}. 
This leads to ambiguity in $a/M$. All claims
to measure or constrain $a/M$ so far are based on the argument that
there is no Fe~K line emission inside the radius of marginal
stability (which varies from
$6r_{g}$ for $a/M=0$ to $1.23 r_{g}$ for $a/M=0.9982$),
and that at $6r_{g}$
one cannot get a large enough gravitational redshift to account
for the strength of the red wing in 
MCG~$-$6$-$30$-$15. Combined with
assumptions about the radial emissivity, this 
can lead one to derive 
near-maximal values of $a/M$ with tiny statistical errors.
However, Figure~4a (solid line) shows that at 
$6r_{g}$, {\it even for a/M=0}, a 6.4~keV line can be redshifted down
to $\sim 3.7$~keV
(for a disk inclination angle of $\theta_{0}=30^{\circ}$;
even lower energies are possible for greater $\theta_{0}$
-- see e.g. Zakharov \& Repin 2006).
Even with the highest signal-to-noise data currently
available it is difficult to tell where the Fe~K line profile
joins the continuum at low energies in a model-independent way.
Thus, a Schwarzschild black hole can
produce a large enough redshift, {\it without invoking line emission
inside the marginally stable radius, or plunge region}.

From a $\sim 290$~ks {\it XMM-Newton} spectrum
for MCG~$-$6$-$30$-$15 Dovciak, Karas, \& Yaqoob 
(2004) first derived
constraints on $a/M$ 
from the counts spectrum
(as opposed to unfolded spectra which 
already assume a value for $a/M$). 
From one of the classes of models fitted,
Figure~4b shows confidence contours 
of $\theta_{0}$ versus $a/M$, and they are flat, covering
all possible values  of $a/M$ from 0 to 1. Details of the model and
parameters can be found in Dov\v{c}iak et al. (2004) but
note that the best-fit inner disk radius was
$(7.0 \pm 0.2)r_{g}$ -- i.e. {\it not} inside the plunge region.
Fe~K line emission inside the plunge region (e.g. Krolik \& Hawley 2002)
may indeed be expected to be
weak (e.g. Brenneman \& Reynolds 2006, and references
therein) but the reduction in line emission with radius
must be quantified. 
To demonstrate why this is important,
Figure~4c shows a ``devil's advocate'' solution to the same data.
It has a Schwarzschild black hole ($a/M=0$), and {\it
no emission inside $6r_{g}$} (the outer radius fitted is $\sim 7r_{g}$).
Obviously this is too simplistic but any radial emissivity
profile that provides sufficient enhancement near $6r_{g}$
will work. This kind of scenario must be shot down first before
measurements of $a/M$ can be placed on a robust footing.
The model includes a complex photoionized absorber and relativistically
blurred Compton reflection consistent with the emission line.

In order to robustly constrain black-hole spin using
{\it Constellation-X} we need to utilize
information that is not {\it all} radially integrated over the
disk and that does not require knowledge
of how the Fe~K line responds to the X-ray continuum.
One can 
constrain $\theta_{0}$
from the persistent, time-averaged line profile.
Then one can probe 
azimuthal, non-axisymmetric enhancements in the line emission by 
time-slicing
and look for events in which the extreme centroids 
(say $E_{-}$ and $E_{+}$) of such hotspots
(located at a radius $r$) could be measured
on their orbit (other scenarios
such as spiral density wave enhancements
also may be relevant -- see Fukumura \& Tsuruta 2004). 
In addition, if the signal-to-noise is
sufficiently high to measure the {\it broad} 
Fe~K$\beta$ line parameters, one
could obtain reasonable constraints on
$E_{0}$. Then one has four observables, two unknowns,
and two equations:
$E_{\pm} = f_{\pm}(r,a/M,\theta_{0},E_{0})$, so that one
could in principle determine $a/M$. Even with one piece of
information missing one could obtain constraints on 
$r$ versus $a/M$.
The method does {\it not} require: (1) any knowledge of the line
radial emissivity function; (2) the 
continuum to behave in a certain way, and (3)
an understanding of how the line emission responds
to continuum variability (it is not reverberation).
Such calculations have been discussed at length in the
literature (e.g. Nayakshin \& Kazanas 2001;
Pech\'{a}cek et al. 2005, and references therein) and
observational evidence from currently available
data for the sought-after non-axisymmetric enhancements is now
accumulating fast (e.g. see Turner et al. 2002, 2006;
Iwasawa, Miniutti, \& Fabian 2004;
Fabian \& Miniutti 2005).
Along with the sources
that show persistent broad Fe~K lines,
the number of AGN exhibiting relativistic effects in
the X-ray spectra is
larger now than it was in the ASCA era, making the case for
{\it Constellation-X} as a probe of strong gravity
more compelling than ever.

\acknowledgements 
This work was partially funded by NASA grants NRA-00-01-LTSA-034,
NNG 04GB78A, and {\it Chandra} grant AR4-5009X. 

\end{document}